\documentclass[epj]{svjour}
\usepackage{hyperref}
\usepackage{graphicx}
\usepackage{amsmath,amssymb}
\usepackage{cite}

\newcommand{\m}{{\rm m}}
\renewcommand{\th}{{\rm th}}
\newcommand{\eff}{{\rm eff}}
\newcommand{\opt}{{\rm opt}}
\newcommand{\el}{{\rm el}}
\newcommand{\grav}{{\rm grav}}
\newcommand{\ggrav}{g_{\rm grav}}
\newcommand{\tor}{{\rm tor}}
\newcommand{\lw}{l_{\rm w}}
\newcommand{\kB}{k_{\rm B}}
\newcommand{\etal}{{\it et al.\, }}
\newcommand{\Pin}{P_{\rm in}}
\newcommand{\kappain}{\kappa_{\rm in}}
\newcommand{\Pcirc}{P_{\rm circ}}
\newcommand{\ncirc}{\bar{n}_{\rm circ}}
\newcommand{\omegacav}{\omega_{\rm cav}}
\newcommand{\omegaL}{\omega_{\rm L}}
\newcommand{\omegam}{\omega_\m}
\newcommand{\Finesse}{\mathcal{F}}
\newcommand{\xzpf}{x_{\rm zpf}}
\newcommand{\Cooperativity}{\mathcal{C}}
\newcommand{\shiki}[1]{Eq.~(\ref{#1})}

\begin{document}
\title{Quantum sensing with milligram scale optomechanical systems}
\author{Yuta~Michimura\inst{1,}\thanks{\email{michimura@granite.phys.s.u-tokyo.ac.jp}} \and Kentaro~Komori\inst{2,}\thanks{\email{kentarok@mit.edu}}}

\institute{Department of Physics, University of Tokyo, Bunkyo, Tokyo 113-0033, Japan \and LIGO Laboratory, Massachusetts Institute of Technology, Cambridge, MA 02139, USA}
\date{\today}
%
\abstract{Probing the boundary between classical and quantum mechanics has been one of the central themes in modern physics. Recently, experiments to precisely measure the force acting on milligram scale oscillators with optical cavities are attracting interest as promising tools to test quantum mechanics, decoherence mechanisms, and gravitational physics. In this paper, we review the present status of experiments using milligram scale optomechanical systems. We compare the feasibility of reaching the quantum regime with a pendulum, torsion pendulum, and optically levitated mirror. Considerations for designing a high $Q$ pendulum, condition for torsion pendulums to have better force sensitivity than pendulums, and constraints in designing optical levitation of a mirror are presented.}
\maketitle
\section{Introduction}
Over the past years, both classical physics and quantum mechanics has been astonishingly successful for explaining macroscopic world and microscopic world, respectively. However, Nature's laws at the interface between classical and quantum mechanics are still not well understood. Experiments to test quantum mechanics at macroscopic scales are naturally the strong driving force of modern physics. This includes the demonstration of quantum superposition states with superconducting quantum interference devices~\cite{SQUID}, Bose-Einstein condensates~\cite{BEC} and complex molecules~\cite{Molecule10000amu,Molecule25kDa}.

With recent progress in cavity optomechanics~\cite{YanbeiReview,AspelmeyerReview}, physicists have demonstrated the preparation of even more macroscopic objects into motional quantum ground state using light~\cite{Chan2011,Teufel2011,Peterson2016}. The progress is especially drastic at scales below nanograms, including recent demonstration of quantum back action measurement at 50~ng~\cite{LSU2019}, but there are also experiments using mechanical oscillators in the range of micrograms to even kilograms~\cite{Kuhn2011,Arcizet2006,Pontin2018,Matsumoto2019,Komori2020,Mow-Lowry2008,Corbitt2007,LIGO2009,LIGO2020}. Most of the effort in these mass ranges is especially focused on reducing the thermal decoherence by reducing the mechanical losses, to bring the oscillators into quantum regimes.

Within these experiments, milligram scale oscillators are drawing attention also as precise gravity sensors~\cite{mgAspelmeyer,Matsumoto2019}, and as possible tools to experimentally explore quantum gravity~\cite{Marletto2017,Bose2017,Belenchia2018,Carney2019}. The other application would be to use these oscillators for testing the speculation that gravity might play a role in destroying macroscopic quantum superposition states, and for testing other wave function collapse models~\cite{DiosiPRL2015,CollapseReview}. Moreover, as a sensitive force sensor, there's a suggestion to use quantum mechanical oscillators for ultralight dark matter detection through non-standard interactions between mirrors and dark matter~\cite{GrahamDM,PierceDM,CarneyDM}.

In this paper, we review the optomechanical experiments using milligram scale oscillators. We start by describing the basic concepts of optomechanical systems and discuss their noise sources for force sensing. We then discuss some of the criteria for reaching the quantum regime with optomechanical systems. Finally, we review the present status of milligram scale experiments with three different types of oscillators: pendulum, torsion pendulum, and optically levitated mirror. We compare these approaches in terms of thermal noise and the feasibility for reaching the quantum regime.

\section{Concepts of optomechanical systems and their force sensitivity}
\begin{figure}
  \begin{center}
    \resizebox{0.9\columnwidth}{!}{\includegraphics{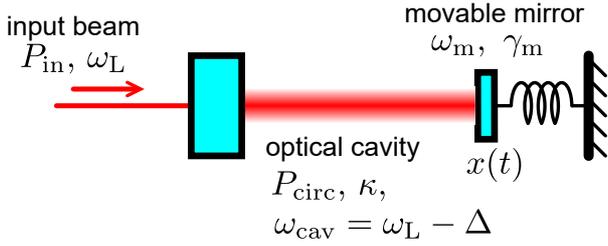}}
  \end{center}
  \caption{Schematic of a simple optomechanical system. The displacement $x(t)$ of the movable mirror with a resonant frequency of $\omegam$ and the energy damping rate of $\gamma_\m$ is being measured. The optical cavity with an amplitude decay rate of $\kappa$ is pumped with a laser light with frequency of $\omegaL$ and input power of $\Pin$. The nearest cavity resonance $\omegacav$ is slightly detuned by $\Delta$, and the circulating power inside is $\Pcirc$.}
  \label{OptomechanicalSystem}
\end{figure}

In optomechanical systems, the displacement of, or the force acting on the mechanical oscillators are read out with light. Radiation pressure interaction between light and mechanical oscillators is often involved in these systems, and optical cavities are used to enhance the interaction. In this section, we briefly describe the basic concepts of mechanical oscillators, optical cavities, and the optomechanical interaction. We then discuss their force sensitivity by describing the quantum noise and the thermal noise. A simple optomechanical system discussed in this section is depicted in Fig.~\ref{OptomechanicalSystem}.

\subsection{Mechanical oscillator}
A mechanical oscillator has multiple vibration modes with resonant frequencies which are determined by the geometry and the material properties. If we focus on a single eigenmode, the displacement of the oscillator $x(t)$ can be well described with the equation of motion of a damped harmonic oscillator:
\begin{equation} \label{EquationofMotion}
 m [ \ddot{x}(t) + \gamma_{\m} \dot{x}(t) + \omegam^2 x(t) ] = F(t) .
\end{equation}
Here, $m$ is the mass and $F(t)$ is the external force. $\omegam$ is the mechanical resonant frequency and $\gamma_\m$ is the energy damping rate, which describes the rate of the loss of mechanical energy. The mechanical quality factor at the mechanical resonance is given by $Q_{\m} = \omegam / \gamma_\m$, and the loss angle is given by $\phi_\m = \gamma_\m \omega / \omegam^2$. We note here that $\gamma_m$ can be frequency dependent, as we will discuss in Sec.~\ref{SECthermalnoise}. Throughout this paper, we will refer to angular frequency as frequency.

The frequency response function from the external force to the displacement can be calculated from the Fourier transform of \shiki{EquationofMotion}, and is called mechanical susceptibility:
\begin{equation}
 \chi_\m (\omega) = \frac{x(\omega)}{F(\omega)} = \frac{1}{m [\omegam^2 - \omega^2 + i \gamma_\m \omega]},
\end{equation}
where $\omega$ is the Fourier frequency. At the mechanical resonance, the magnitude of the susceptibility is enhanced by a factor of $Q_\m$, compared with the low frequency limit $ \left. \chi_\m \right|_{\omega \ll \omegam} = 1/(m \omegam^2)$. At the frequencies well above the mechanical resonance, the oscillator behaves as free-mass and $ \left. \chi_\m(\omega) \right|_{\omega \gg \omegam} = -1/(m \omega^2)$.

One of the necessary conditions to bring a mechanical oscillator to quantum regime is to cool down the oscillator to its mechanical ground state. When the oscillator is coupled to a high-temperature bath of $T_\th$, the average phonon number of the oscillator $\bar{n}$ equals to $\bar{n}_\th \simeq \kB T_\th / (\hbar \omegam) $. When $\bar{n}<1$ is achieved, we refer to as ground state cooling is achieved since the quantum zero-point fluctuation of the oscillator can be observed. At the ground state $\bar{n}=0$, an oscillator has the zero-point energy $\hbar \omegam/2$.  The time dependence of the phonon number for an oscillator that is initially cooled down to the ground state can be expressed with
\begin{equation}
  \bar{n} = \bar{n}_\th (1 - e^{-\gamma_\m t}) .
\end{equation}
At $t=0$, the phonon number increases with
\begin{equation} \label{Eq:ThermalDecoherenceRate}
 \Gamma_\th \equiv \frac{{\rm d} \bar{n}}{{\rm d} t} = \bar{n}_\th \gamma_\m \simeq \frac{\kB T_\th}{\hbar Q_\m} .
\end{equation}
This $\Gamma_\th$ is referred to as the thermal decoherence rate, and this implies that achieving a high mechanical $Q$ factor and a low-temperature bath are important for low thermal decoherence and low thermal noise.

\subsection{Optical cavities}
An optical cavity is an arrangement of mirrors which creates the round-trip of light propagation. The light resonates inside the cavity when the round-trip length is integer multiples of the wavelength of light $\lambda$. Here, we consider a simple Fabry-P\'{e}rot cavity consisting of two highly reflective mirrors separated by a distance $L$, pumped with a single frequency continuous-wave laser source.

The laser beam resonates inside the cavity when the laser frequency $\omegaL=2 \pi c / \lambda$ is integer multiples of the free spectral range given by
\begin{equation} \label{FSR}
  \omega_{\rm FSR} = \frac{\pi c}{L} .
\end{equation}
We define the difference between the laser frequency and its nearest cavity resonant frequency as the laser detuning $\Delta = \omegaL - \omegacav$. The ratio of the free spectral range to the full-width half-maximum of the resonance is called finesse, and is given by
\begin{equation}
 \Finesse = \frac{\omega_{\rm FSR}}{2 \kappa},
\end{equation}
where $\kappa$ is an amplitude decay rate of the cavity, also known as the cavity pole frequency. The circulating power inside the cavity when pumped with the input power of $\Pin$ is
\begin{equation} \label{EqPcirc}
 \Pcirc = \frac{2 \Finesse}{\pi} \frac{\kappain}{\kappa} \frac{\Pin}{1+\left(\Delta/\kappa\right)^2} .
\end{equation}
Here, $\kappain$ is the amplitude decay rate associated with the input coupling. When the input mirror has the intensity transmission of $T_{\rm in}$, $\kappain$ is given by
\begin{equation}
 \kappain = \frac{T_{\rm in}c}{4 L} .
\end{equation}
The ratio $\kappain/\kappa$ determines the circulating power, and when $\kappain/\kappa = 1/2$, the cavity is called a critically coupled cavity. When $\kappain/\kappa$ is larger (smaller) than $1/2$, the cavity is called an over-coupled (under-coupled) cavity. For over-coupled cavity, most of the beam is reflected back towards the laser source. When $\kappa - \kappain$ is constant, which means that all the optical losses other than $\kappain$ is constant, critically-coupled cavity gives the largest circulating power. The factor $2 \Finesse / \pi$ gives the average number of round-trips before a photon leaves the cavity.

\subsection{Optical spring and optomechanical cooling} \label{Sec:OpticalSpring}
As apparent from \shiki{FSR}, the motion of the mirror $\delta x$ changes the cavity resonant frequency $\omegacav$ by 
\begin{equation} \label{OpticalFrequencyShift}
  \delta \omegacav = \frac{\omegacav}{L} \delta x \equiv G \delta x ,
\end{equation}
since $\omegacav$ is an integer multiple of $\omega_{\rm FSR}$. Therefore, the motion of the mirror changes the cavity detuning. The circulating power inside the optical cavity is proportional to the position of the mirror, when the mirror motion is small and the cavity is detuned from the laser frequency, as indicated by \shiki{EqPcirc}. Therefore, the radiation pressure acts as an optical spring to the mirror. This optical spring has a time delay since it takes a finite time to change the circulating power, and optomechanical damping is involved. We can use this optical spring and optomechanical damping to manipulate the susceptibility of the mirror, and to cool the motion of the mirror.

The resonant frequency shift $\delta \omega_\opt$ from the optical spring and the optomechanical damping rate $\gamma_\opt$ is given by~\cite{AspelmeyerReview}
\begin{eqnarray}
 \delta \omega_\opt &=& g^2 \left( \frac{\Delta + \omegam}{\kappa^2+ (\Delta + \omegam)^2} + \frac{\Delta - \omegam}{\kappa^2+ (\Delta - \omegam)^2} \right) , \label{Eq:OpticalSpring} \\
 \gamma_\opt &=& g^2 \left( \frac{2\kappa}{\kappa^2+ (\Delta + \omegam)^2} - \frac{2\kappa}{\kappa^2+ (\Delta - \omegam)^2} \right) , \label{OpticalDamping}
\end{eqnarray}
where $g$ is the optomechanical coupling strength given by
\begin{equation}
 g^2 = \frac{\Pcirc \omegaL}{m L c \omegam} .
\end{equation}
This optomechanical coupling strength can be rewritten with
\begin{equation}
 g = G \xzpf \sqrt{\ncirc} ,
\end{equation}
where $G$ is the optical frequency shift per oscillator displacement defined in \shiki{OpticalFrequencyShift}, $\xzpf$ is the zero point fluctuation of the oscillator given by $\xzpf = \sqrt{\hbar/(2 m \omegam)}$, and $\ncirc$ is the average number of photons circulating inside the cavity given by
\begin{equation}
 \ncirc = \frac{2L}{c} \frac{\Pcirc}{\hbar \omegaL} .
\end{equation}

In pendulum or torsion pendulum experiments, $\omegam /(2 \pi)$ is usually below $\sim 1$~Hz, and linewidth of the cavity is larger than the mechanical resonant frequency, i.e. $\kappa \gg \omegam$. This regime is referred to as the Doppler regime, or the bad cavity regime. Under this regime, Eqs.~(\ref{Eq:OpticalSpring}) and (\ref{OpticalDamping}) yields
\begin{eqnarray}
 \left. \delta \omega_\opt \right|_{\kappa \gg \omegam} &\simeq& g^2 \frac{2 \Delta}{\kappa^2+\Delta^2} , \\
 \left. \gamma_\opt \right|_{\kappa \gg \omegam} &\simeq& -g^2 \frac{8\kappa \Delta \omegam}{(\kappa^2+ \Delta^2)^2} .
\end{eqnarray}

\begin{figure}
  \begin{center}
    \resizebox{0.9\columnwidth}{!}{\includegraphics{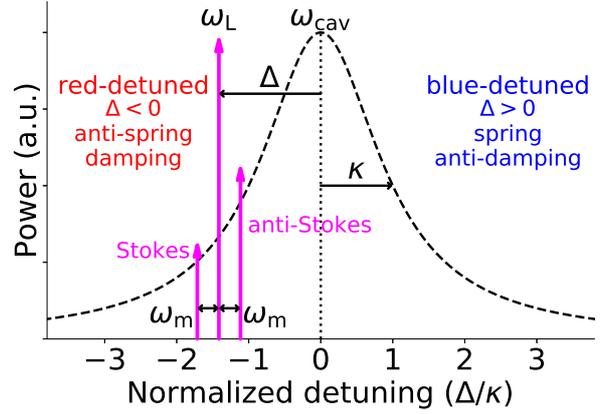}}
  \end{center}
  \caption{Red-detuned laser beam gives optical anti-spring since the circulating power inside the cavity increases with cavity length increase. It also gives optical damping since anti-Stokes sideband is enhanced more than Stokes sideband. Vice versa for blue-detuned laser beam.}
  \label{OpticalSpring}
\end{figure}

The susceptibility of the mirror under optomechanical coupling is
\begin{equation}
 \chi_\eff (\omega) = \frac{1}{m [\omega_\eff^2 - \omega^2 + i \gamma_\eff \omega]},
\end{equation}
with $\omega_\eff = \omegam + \delta \omega_\opt$ and $\gamma_\eff=\gamma_\m +\gamma_\opt$. As indicated from the equations above, red-detuned laser beam ($\Delta < 0$) gives optical anti-spring and damping, while blue-detuned laser beam ($\Delta > 0$) gives optical spring and anti-damping (see Fig.~\ref{OpticalSpring}). $\omega_\eff > 0$ and $\gamma_\eff > 0$ are required for the stability of the system, and the combination of the use of red-detuned laser beam and blue-detuned laser beam are necessary when mechanical spring or mechanical damping is small. This scheme is called double optical spring~\cite{CorbittSpring,DoubleOpticalSpring}, and is often used in pendulum experiments.

\subsection{Force sensitivity}
\begin{figure}
  \begin{center}
    \resizebox{0.9\columnwidth}{!}{\includegraphics{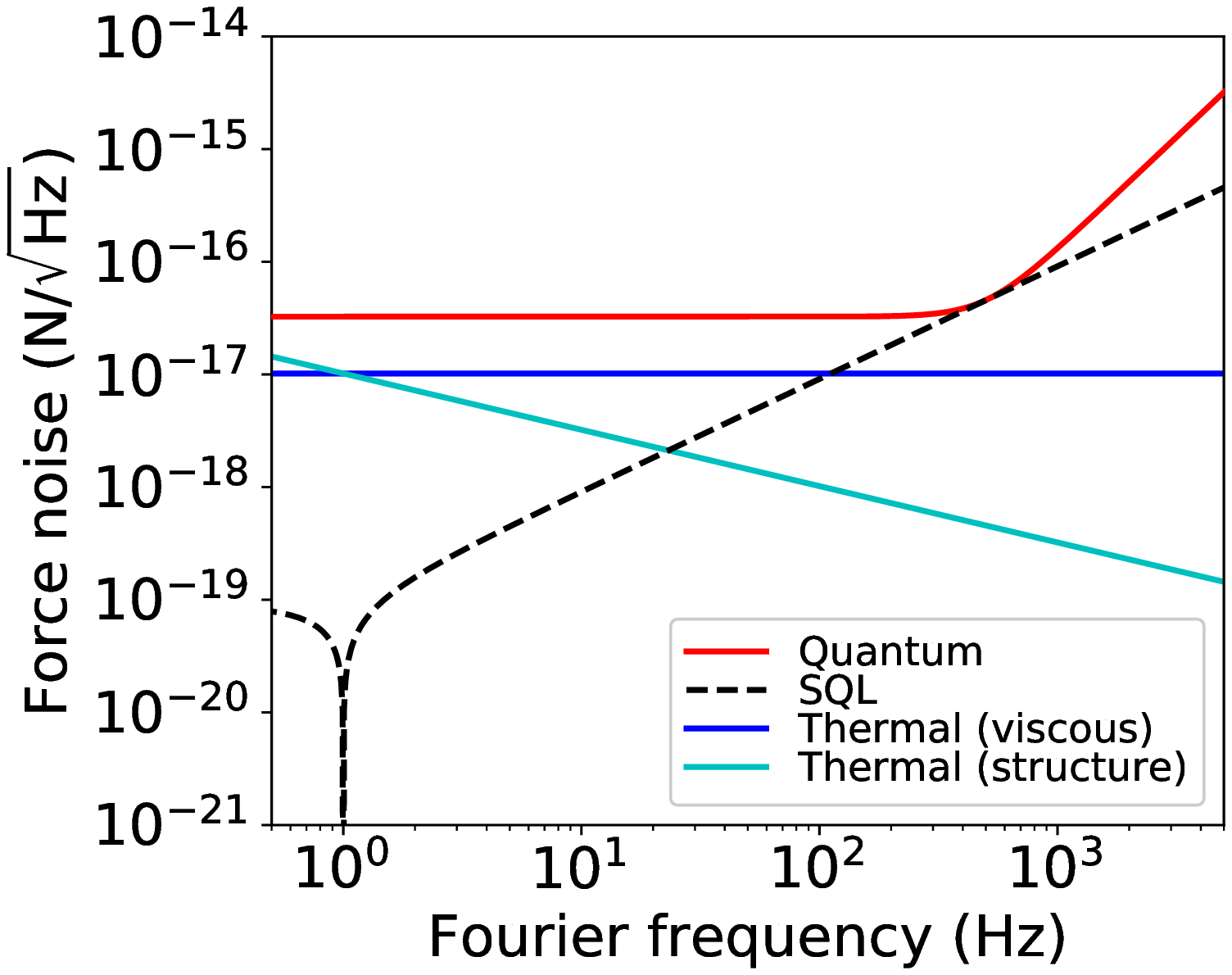}} \\
    \resizebox{0.9\columnwidth}{!}{\includegraphics{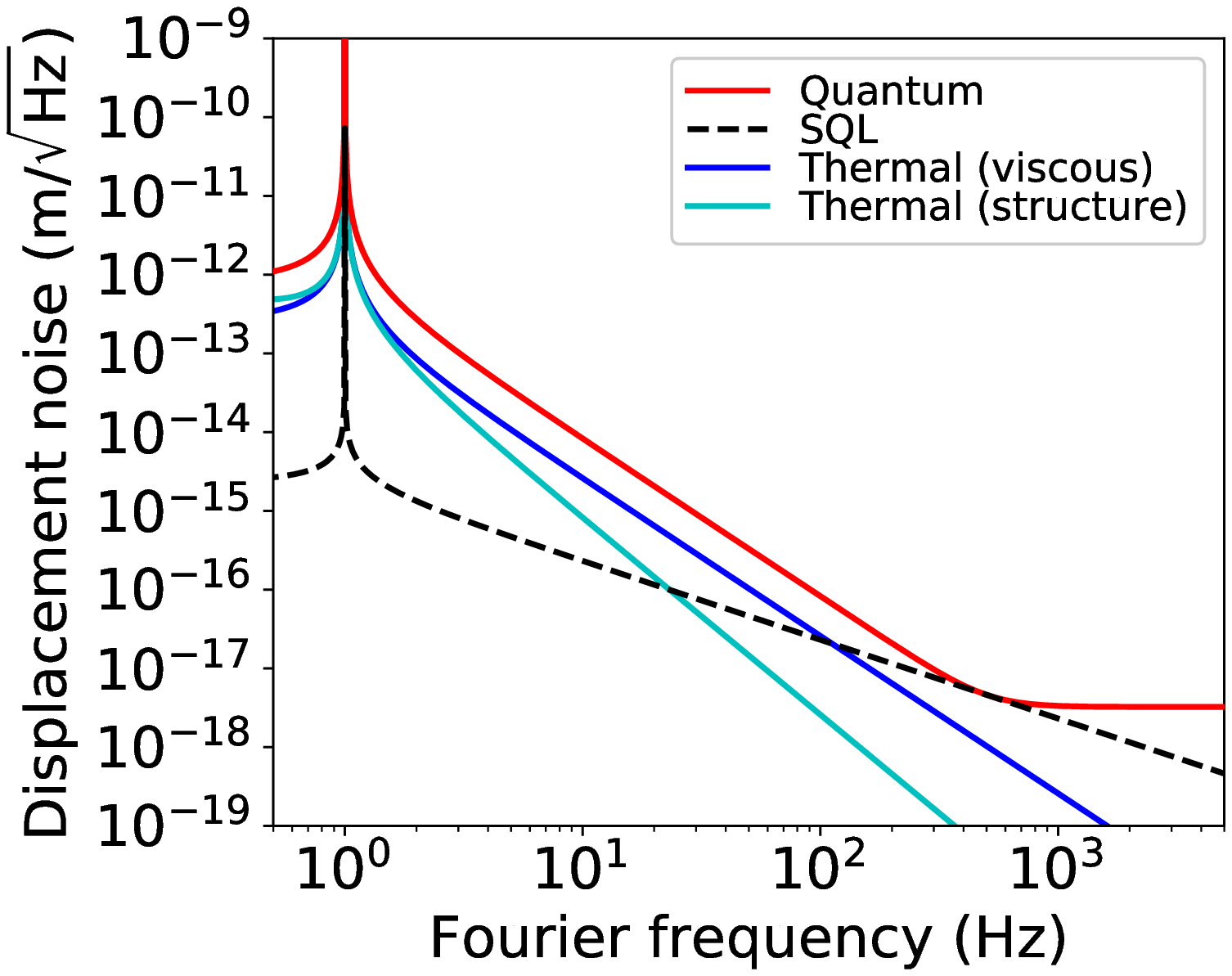}}
  \end{center}
  \caption{Quantum noise and thermal noise spectrum in force (top) and displacement (bottom) with example parameters. Thermal noise for viscous and structure dissipation models at room temperature are plotted. Mechanical oscillator with $\omegam=2 \pi \times 1$~Hz, $Q_\m=10^9$, $m=1$~mg, $T_\th=300$K, and optical cavity with $\Finesse=10^2$, $L=10$~cm, $\Pcirc=1$~W, $\lambda=1064$~nm is assumed. The standard quantum limit touching frequency is $\omega_{\rm SQL} \simeq 2 \pi \times 500$~Hz.}
  \label{ExampleSensitivity}
\end{figure}

The force sensing of the mechanical oscillator with an optical cavity is ultimately limited by the quantum fluctuation of light and the thermal fluctuation of the oscillator. The quantum noise and the thermal noise associated with the mechanical mode of interest is described below. We note here that there are multitude of other classical noises to be reduced for reaching the ultimate sensitivity, such as laser intensity noise, laser frequency noise, seismic noise, and other technical noises. For discussion on such classical noises, the reader is referred to, for example, the review article by Adhikari~\cite{RanaReview}. Force and displacement sensitivity spectra with example parameters are shown in Fig.~\ref{ExampleSensitivity}.

\subsubsection{Quantum noise}
If we assume the cavity detuning $\Delta=0$ for simplicity, the single-sided power spectral density of the quantum noise for an optical cavity in displacement is given by~\cite{YanbeiReview}
\begin{equation} \label{quantumnoise}
 S_{\rm qn}^{x}(\omega) = \frac{x_{\rm SQL}^2}{2} \left( \frac{1}{\mathcal{K}} + \mathcal{K} \right),
\end{equation}
with
\begin{equation}
 x_{\rm SQL} = \sqrt{2 \hbar | \chi_{\rm M}(\omega) |}
\end{equation}
and
\begin{eqnarray}
 \mathcal{K} &=& \frac{8 \omegaL \Pcirc |\chi_{\rm M}(\omega)|}{Lc \kappa} \frac{1}{1+\left( \omega/\kappa \right)^2} \\
  &=& \frac{4 \hbar G^2 \ncirc |\chi_{\rm M}(\omega)|}{\kappa} \frac{1}{1+\left( \omega/\kappa \right)^2} .
\end{eqnarray}
Here, $\chi_{\rm M}(\omega)$ is the susceptibility for the reduced mass of the system. For a Fabry-P{\'e}rot cavity constructed from a movable mirror of a mass $m$ and an input mirror much heavier than $m$, the reduced mass is $M=m$. For a Michelson interferometer with two of such identical Fabry-P{\'e}rot cavities who's intra-cavity power is $\Pcirc/2$ each in both arms, the reduced mass is $M = m / 2$. We note that in papers related to gravitational wave detectors, such as Refs.~\cite{Kimble2001,BuonannoChen}, single-sided spectral density in differential measurement of two cavities are often presented, while in other papers discussing cavity optomechanics such as Ref.~\cite{AspelmeyerReview}, double-sided spectral density in one cavity is presented.

The first term in \shiki{quantumnoise} represents the shot noise, which decreases with $P_{\rm circ}$, and the second term represents the quantum radiation pressure noise, or the back action noise, which increases with $P_{\rm circ}$. Because of this trade-off betwen the shot noise and the quantum radiation pressure noise, $S_{\rm qn}^{x}(\omega) \ge x_{\rm SQL}^2$, and this limit is called the standard quantum limit.

For a system with $M=m$, the quantum radiation pressure noise is
\begin{equation}
 S_{\rm rad}^{x}(\omega) = |\chi_\m(\omega)|^2 \frac{4 \hbar^2 G^2 \ncirc}{\kappa} \frac{1}{1+\left( \omega/\kappa \right)^2} ,
\end{equation}
and the shot noise is
\begin{equation} \label{Eq:ShotNoise}
 S_{\rm shot}^{x}(\omega) = \frac{\kappa}{4 G^2 \ncirc} \left[ {1+\left( \omega/\kappa \right)^2} \right] .
\end{equation}
When there are optical losses in the displacement detection, and the photon collection efficiency $\eta$ is smaller than unity, the shot noise will be $S_{\rm shot}^{x}(\omega) / \eta$. The quantum noise in force can be calculated with $S_{\rm qn}^{F}=S_{\rm qn}^{x}/|\chi_\m(\omega)|^2$.

The standard quantum limit is reached when the quantum radiation pressure noise equals to the shot noise. When reached at free mass region in the Doppler regime, i.e. $\omegam \ll \omega \ll \kappa$, the standard quantum limit touching frequency will be
\begin{equation}
 \omega_{\rm SQL} = \sqrt{\frac{S_{\rm rad}^{F}}{\hbar m}} = \sqrt{\frac{4 \hbar G^2 \ncirc}{m \kappa}} . \label{Eq:SQLtouch}
\end{equation}

\subsubsection{Thermal noise} \label{SECthermalnoise}
Thermal noise can be calculated by applying the fluctuation-dissipation theorem to the damped harmonic oscillator, and its single-sided power spectral density in force is simply given by~\cite{Saulson1990}
\begin{equation}
 S_\th^{F}(\omega) = 4 \kB T_{\th} m \gamma_\m .
\end{equation}

There are two dissipation models, namely viscous model and structure model. In the viscous model, the oscillation is damped by a force proportional to the velocity, and the damping rate is frequency independent. In the structure model, the energy loss comes from the internal loss of the mechanical oscillator, and the loss angle is frequency independent. The complex spring constant in the structure model can be written as $k_\m (1+i \phi_\m)$. The energy damping rate in each model is given by
\begin{eqnarray}
 \gamma_\m^{\rm viscous} &=& \frac{\omegam}{Q_\m} , \label{Eq:Viscous} \\
 \gamma_\m^{\rm structure} (\omega) &=& \frac{\omegam^2}{\omega Q_\m} . \label{Eq:Structure}
\end{eqnarray}
This suggests that, for the same mechanical quality factor at the mechanical resonance, structure model gives smaller noise at frequencies above the mechanical resonance (see Fig.~\ref{ExampleSensitivity}). For a 1~mg oscillator with $\omegam=2 \pi \times 1$~Hz, $Q_\m$ of $10^5$ is required to achieve $1~{\rm fN/\sqrt{Hz}}$ force sensitivity at room temperature.

In this paper, we focus on the thermal noise from the dissipation of the main oscillator mode of interest. However, in actual experiments, other modes of the oscillator will couple to the measurement. For experiments using a mirror as an oscillator, thermal noise from the high reflective coating and mirror substrate bulk modes are especially important in designing the experiment. There are thermal fluctuations in the coatings from thermal dissipation, which also produce noise from the thermoelastic and thermorefractive mechanisms. Such thermal noises has been studied extensively in the field of interferometric gravitational wave detectors, and the reader is referred to other articles for a more complete discussion~\cite{LevinSubstrate,HarryCoating,CerdonioThermoelastic,SomiyaThermoelastic,EvansThermooptic}.

\section{Criteria for reaching the quantum regime} \label{Sec:Criteria}
There is a number of criteria to optomechanical systems for reaching the quantum regime. For example, the experiment proposed in Ref,~\cite{Marshall2003} to create quantum superposition states of a mirror via the interaction with a single photon require motional ground state cooling of the mirror and ultra-strong optomechanical coupling. The experiment proposed in Ref.~\cite{MullerEbhardt2008} to create entanglement between the motions of two mirrors require reaching the standard quantum limit of measuring the common and differential displacement of two mirrors. The experiment to see quantum correlation of light mediated by gravitational interaction between two optomechanical cavities require quantum radiation pressure noise limited sensitivity~\cite{Haixing2019}.

Here, we review three criteria especially important for discussing milligram scale optomechanical systems. We first describe the criterion for quantum radiation pressure noise limited measurement. We then describe the criteria for the measurement rate and $f \cdot Q$ for achieving ground state cooling of mechanical oscillators.

\subsection{Quantum radiation pressure noise limited measurement} \label{Sec:QRPNlimited}
When the quantum radiation pressure noise is the dominating force noise, this means that the quantum fluctuation of light is transferred to the oscillator. It is also required that all the classical force noise is smaller than the quantum radiation pressure noise up to $\omega_{\rm SQL}$ to reach the standard quantum limit in the free-mass region, which is the necessary condition to prepare a conditional-state~\cite{YanbeiReview}.

Assuming thermal noise is the dominant classical force noise, this requirement is reduced to $S_{\rm rad}^{F} / S_\th^{F} > 1$. For $\omega \ll \kappa$,
\begin{eqnarray}
 \frac{ S_{\rm rad}^{F} } { S_\th^{F} } &=& \frac{ 4 \hbar^2 G^2 \ncirc }{\kappa} \frac{1}{4 \kB T_{\th} m \gamma_\m} \nonumber \\
  &=& \frac{2 g^2}{\gamma_\m \kappa} \frac{1}{\bar{n}_\th} \\
  & \equiv & \frac{\Cooperativity}{\bar{n}_\th} \equiv \Cooperativity_{\rm qu} .
\end{eqnarray}
Here, $\Cooperativity$ is called cooperativity, and $\Cooperativity_{\rm qu}$ is called quantum cooperativity~\cite{AspelmeyerReview}. $\Cooperativity_{\rm qu} > 1$ is the requirement for assuring that the state transfer between light and oscillator is faster than the mechanical decoherence rate, and is equivalent to making the quantum radiation pressure noise larger than the thermal noise.

\subsection{Measurement rate}
To cool the motion of the mechanical oscillator towards its ground state, optomechanical cooling explained in Sec.~\ref{Sec:OpticalSpring} can be used. Optomechanical cooling is especially effective in the resolved sideband regime ($\kappa \ll \omegam$), and ground state cooling is actually possible without any additional damping. In the Doppler regime ($\kappa \gg \omegam$) however, Stokes sideband and anti-Stokes sideband is not resolved, and optomechanical cooling is not very effective. To increase the optomechanical damping rate $\gamma_\opt$, circulating power inside the cavity needs to be increased, which will result in more quantum radiation pressure noise. In the end, minimum phonon number we can obtain will be~\cite{AspelmeyerReview},
\begin{equation}
 \left. \bar{n}_{\rm min} \right|_{\kappa \gg \omegam} = \frac{\kappa}{2 \omega_\m} \gg 1 ,
\end{equation}
and additional damping scheme is required for ground state cooling.

A conventional way to cool the oscillator in the Doppler regime is to apply a feedback force proportional to the velocity of the oscillator~\cite{FeedbackCooling}. This scheme is called feedback cooling, or cold damping. The estimation of the velocity is done by taking the time derivative of the measured displacement, and therefore the measurement imprecision will be crucial in this scheme. The ratio of the zero point variance of the oscillator to the measurement imprecision is called the measurement rate, and is given by~\cite{MeasurementRate,Wilson2015}
\begin{equation}
  \Gamma_{\rm meas} \equiv \frac{\xzpf^2}{2 S_{\rm imp}^x(\omega_\m)} ,
\end{equation}
where $S_{\rm imp}^x$ is the sensing noise spectrum calibrated into the oscillator displacement. The measurement rate can be understood as the inverse of the time it takes to measure the zero point fluctuation. The measurement rate has to be large enough compared with the thermal decoherence rate defined in \shiki{Eq:ThermalDecoherenceRate} for ground state cooling, i.e.
\begin{equation} \label{Eq:MeasurementRateRequirement}
  \Gamma_{\rm meas} \gtrsim \frac{\Gamma_\th}{8} .
\end{equation}

When the sensing noise equals to the shot noise, which is an ultimate limit to the displacement sensing, measurement rate reduces to
\begin{eqnarray}
  \left. \Gamma_{\rm meas} \right|_{S_{\rm imp}^x = S_{\rm shot}^x} &=& \frac{2 g^2}{\kappa} \nonumber \\
  &=& \Cooperativity \gamma_\m ,
\end{eqnarray}
with $\omega \ll \kappa$. Therefore, the criterion for the measurement rate in \shiki{Eq:MeasurementRateRequirement} reduces to $\Cooperativity_{\rm qu} \gtrsim 1/8$. This criterion can be met if $S_{\rm rad}^{F} / S_\th^{F} > 1$, as discussed in Sec.~\ref{Sec:QRPNlimited}.

\subsection{$f \cdot Q$ criterion}
Another criterion for ground state cooling usually discussed in the literature is $f \cdot Q$ criterion. The phonon number of the oscillator after cooling can be written as
\begin{equation}
 \bar{n}_\eff = \frac{\kB T_\eff}{\hbar \omega_\eff} = \frac{\kB T_\th}{\hbar \omega_\eff} \frac{\gamma_\m(\omega_\eff)}{\gamma_\eff(\omega_\eff)},
\end{equation}
where $T_{\eff}$ is the effective temperature of the eigenmode. To avoid over-damping, $\gamma_\eff < \omega_\eff$ is required and we get
\begin{equation}
 \bar{n}_\eff > \frac{\kB T_\th}{\hbar \omega_\eff Q_\eff} ,
\end{equation}
with $Q_\eff \equiv \omega_\eff / \gamma_\m (\omega_\eff)$. To have $\bar{n}_\eff < 1$, right hand side of the above inequality have to be much smaller than unity, which reduces to
\begin{equation}
 f_\eff Q_\eff \gg \frac{\kB T_\th}{h} \simeq 6 \times 10^{12} \left( \frac{T_\th}{300~\rm{K}} \right) ,
\end{equation}
where $f_\eff = \omega_\eff/(2 \pi)$.

If we use $f_\m Q_\m$ instead of $f_\eff Q_\eff$, this criterion becomes
\begin{equation}
 f_\m Q_\m \gg 6 \times 10^{12} \left( \frac{T_\th}{300~\rm{K}} \right) \left( \frac{\omega_\m}{\omega_\eff} \right)^\alpha ,
\end{equation}
where $\alpha=2$ if the mechanical damping follows the viscous damping model, which has frequency independent damping rate given in \shiki{Eq:Viscous}, and $\alpha=3$ if it follows the strusture damping model, which has frequency dependent damping rate given in \shiki{Eq:Structure}. This suggests that $f \cdot Q$ criterion can be relaxed by increasing the eigenmode frequency with optical spring. This is because mechanical loss associated with the mechanical spring is diluted by optical spring free of dissipation. Such optical dilution is more effective in structure damping, since $\gamma_\m$ is reduced with frequency.

\section{Milligram scale experiments}
So far, we have discussed the basics of optomechanical systems without assumptions in the type of the mechanical oscillators. At macroscopic scales, pendulum modes of suspended mirrors are often used since they can have a very high mechanical $Q$ factor. In this section, we discuss practical considerations to realize optomechanical experiments using such pendulums. We start by describing dissipation mechanisms in pendulums and radiation pressure induced angular instability of the optomechanical cavities. Current status of pendulum experiments at milligram scales are also presented as needed. An alternative approach for realizing quantum sensing with a suspended mirror is to use a torsional mode of a mirror. We make a comparison between torsion pendulum and pendulum in terms of quantum cooperativity and thermal noise. Finally, we discuss the use of radiation pressure alone to support a mirror to realize optical levitation, instead of supporting a mirror with lossy mechanical structures. Comparison of experiments discussed in this section is summarized in Table~\ref{tab:experiments}.

\subsection{Dissipation mechanisms in pendulums}
In a pendulum, the most of the restoring force comes from the gravitational field, and a small amount of the restoring force comes from intrinsic elasticity of the wire suspending the mass. The gravitational spring is dissipation free, while the intrinsic spring follows the structure damping model given in \shiki{Eq:Structure}. The total mechanical complex spring constant is given by
\begin{equation}
 K_{\rm m}^{\rm pend} = k_{\grav} + k_{\el}(1 + i \phi_{\el}) \equiv k_{\rm m}^{\rm pend} \left( 1 + i \phi_{\rm m}^{\rm pend} \right) ,
\end{equation}
where $k_{\grav}$ and $k_{\el}$ are the spring constants of the gravitational restoring force and the intrinsic elasticity, respectively. $\phi_{\el}=1/Q_{\el}$ is the loss angle of the intrinsic elasticity. For a pendulum with a thin wire, $k_{\grav}$ is much larger than $k_{\el}$, and the mechanical $Q$ of the pendulum mode will be
\begin{equation}
 Q_{\rm m}^{\rm pend} \simeq \frac{k_{\grav}}{k_{\el}} Q_{\el} \equiv \Lambda Q_{\el} .
\end{equation}
$1/\Lambda$ is called a dissipation dilution factor, and the mechanical $Q$ factor of a pendulum can be much larger than the intrinsic $Q$ of the wire suspending the mass by a factor of $\Lambda$. The gravitational spring and the elastic spring for a pendulum of length $\lw$ are given by
\begin{eqnarray}
 k_{\grav} &=& \frac{m \ggrav}{\lw}, \\
 k_{\el} &=& \frac{n_{\rm w} \sqrt{\pi T_{\rm w} E_{\rm w} I_{\rm w}} }{2 \lw^2} ,
\end{eqnarray}
respectively~\cite{Saulson1990}. Here, $\ggrav$, $n_{\rm w}$, $T_{\rm w}$, and $E_{\rm w}$ are the gravitational acceleration, the number of wires suspending the mass, the tension in each wire, and the Young's modulus, respectively. For a cylindrical wire with a radius of $r_{\rm w}$, the moment of inertia of the wire cross section $I_{\rm w}=\pi r_{\rm w}^4/4$. When all the suspension wires are vertical, the tension $T_{\rm w}=m \ggrav/n_{\rm w}$, and the $Q$ enhancement factor will be
\begin{equation}
 \Lambda = \frac{4 \lw}{r_{\rm w}^2} \sqrt{\frac{m \ggrav}{\pi n_{\rm w} E_{\rm w}}} .
\end{equation}

This equation clearly shows that it is essential to use a long and thin wire to achieve a high $Q$ pendulum. However, a long pendulum will give lower violin mode frequencies, and violin modes will contaminate the pendulum mode measurement. The first violin mode frequency is given by~\cite{Gaby1994}
\begin{equation}
 f_{\rm v}=\frac{1}{2 l_{\rm w}} \sqrt{\frac{T_{\rm w}}{\rho_{\rm w}\pi r_{\rm w}^2}} ,
\end{equation}
where $\rho_{\rm w}$ is the density of the wire, and we can treat the pendulum mode as a simple harmonic oscillator only well below this frequency. Considering the standard quantum limit touching frequency given in \shiki{Eq:SQLtouch}, it is usually enough if $f_{\rm v}$ is higher than a few kilohertz in laboratory scale experiments.

There is also a limit for thinner wire since the tension of a wire has to be smaller than the wire breaking strength. Using a tensile strength $H_{\rm w}$, this condition reduces to
\begin{equation}
 \pi r_{\rm w}^2 H_{\rm w} = s_{\rm w} T_{\rm w} , \label{Eq:tensile}
\end{equation}
where $s_{\rm w}$ is a safety factor above unity. With these boundary conditions, $Q$ factor of the pendulum mode is given by
\begin{equation}
 Q_{\rm m}^{\rm pend} = \frac{2 H_{\rm w}}{s_{\rm w} f_{\rm v}} \sqrt{\frac{1}{\rho_{\rm w} E_{\rm w}}} \frac{Q_{\el}}{r_{\rm w}} .
\end{equation}
Therefore, a wire with high tensile strength and low density gives a high $Q$ pendulum. For a thin wire, intrinsic mechanical loss is dominated by a surface loss, and the amount of loss is proportional to the surface over the volume~\cite{Penn2006,Numata2007}, in other words, $Q_{\el}/r_{\rm w}$ is constant. Increasing the number of wires $n_{\rm w}$ is generally not effective for increasing the $Q$ factor, and it would introduce other resonant modes. There is also a possiblity to increase $T_{\rm w}$ by suspending the mass with wires from the different directions, but it is also not effective from the discussion above.

The other important dissipation mechanisms in practical pendulums are the losses associated with the clamping and bonding of the wires to the suspension point and the mass, thermoelastic damping, and gas damping. Bonding loss and clamping loss are considered to follow the structure damping model, and it is often difficult to distinguish them from the wire's intrinsic loss. Generally, bonding loss can be reduced by making the volume of the bonding region small to make the stress stored in the region small. In previous milligram to gram scale experiments, epoxy glue has been conveniently used to attach wires to the mirrors, but it is known to have a large loss, with a loss angle in the order of $10^{-2}$~\cite{Neben2012}.

In ground-based gravitational wave detectors such as Advanced LIGO and Advanced Virgo, fused silica fibers are welded to 40--42~kg fused silica test masses to make monolithic suspensions for reducing clamping and bonding losses, and their natural mechanical $Q$ factors for the pendulum mode are estimated to be in the order of up to $10^9$~\cite{VirgoMonolithic,aLIGOMonolithic,ApMonolithic}. KAGRA detector takes a different approach, and uses sapphire fibers to suspend sapphire mirrors for cryogenic cooling. Sapphire fibers with nail heads are hooked on sapphire ears with gallium, and ears are bonded with hydroxide catalysis bonding on both sides of the mirror~\cite{KAGRACryopayload}. Extracting heat through fibers suspending the mass require temperature gradient, and calculation of thermal noise under such a condition is studied in Ref.~\cite{Komori2018}.

Very recently, Cata{\~n}o-Lopez \etal reported the development of 7~mg monolithic fused silica pendulum~\cite{Seth2019}. In their previous setup, they have used a 1~cm long fused silica fiber bonded to a fused silica mirror by epoxy glue, and reported the measured $Q$ value for the pendulum mode at 4.4~Hz to be $1 \times 10^5$~\cite{Matsumoto2019}. For the monolithic pendulum, they have used 5~cm fused silica fiber, and reported the measured $Q$ value of $2 \times 10^6$ at 2.2~Hz. They claim that $f \cdot Q$ criterion can be met by realizing $f_\eff=280$~Hz and $Q_\eff=3 \times 10^{10}$ with optical trapping of this monolithic pendulum.

Thermoelastic damping comes from the relaxation process of the temperature gradient caused by inhomogeneous elastic deformation of the wire~\cite{Zener1,Zener2}. The pendulum motion creates wire bending around clamping and bonding region, and this creates temperature gradient inside the wire. The temperature gradient couples to the mechanical mode of the wire through the thermal expansion. The loss angle for thermoelastic damping is given by
\begin{equation}
 \phi(\omega) = \frac{\Delta_{\rm r} \omega \tau_{\rm r}}{1+(\omega \tau_{\rm r})^2},
\end{equation}
where $\Delta_{\rm r}$ and $\tau_{\rm r}$ are the relaxation strength and the relaxation time, respectively, given by
\begin{eqnarray}
 \Delta_{\rm r} &=& \frac{E_{\rm w} \alpha_{\rm w}^2 T_{\rm th}}{\rho C_{\rm w}} \\
 \tau_{\rm r} &=& \frac{1}{2 \pi} \frac{\rho_{\rm w} C_{\rm w} r_{\rm w}^2}{0.539\kappa_{\rm w}}.
\end{eqnarray}
Here, $\alpha_{\rm w}$, $C_{\rm w}$ and $\kappa_{\rm w}$ are the thermal expansion coefficient, the specific heat capacity and the thermal conductivity, respectively. Thermoelastic damping also reduces by the use of thinner wires, since the relaxation time is shortened. Note that thermoelastic damping can be nulled by applying a suitable amount of static stress to the wire, when the Young's modulus increases with temperature~\cite{Cagnoli2002}.

The damping from residual gas is known to follow the viscous model in \shiki{Eq:Viscous}. The energy damping rate from residual gas is given by~\cite{Saulson1990}
\begin{equation}
 \gamma_\m^{\rm gas} = \frac{p A}{C m} \sqrt{\frac{m_{\rm gas}}{\kB T_\th}} ,
\end{equation}
where $p$ is the pressure, $A$ is the surface area of the mirror, $C$ is a dimension less constant of order unity that depends on the shape of the mirror, and $m_{\rm gas}$ is the mass of one gas molecule. For a 1~mg disk with a diameter of 2~mm at room temperature, $\gamma_\m^{\rm gas} \simeq 10^{-8}$~Hz, when $p=10^{-5}$~Pa and the residual gas is dominated by helium. This gives $Q_\m$ in the order of $10^{8}$, when mechanical resonant frequency is at around 1~Hz. This residual gas damping will likely to limit the $Q$ factor of pendulums, when mechanical dissipations discussed above are minimized by a high quality suspension in the future.

\subsection{Angular instability of optomechanical cavity} \label{SEC:Instability}
In optomechanical experiments, the longitudinal motion of the oscillator can be trapped with optical springs described in Sec.~\ref{Sec:OpticalSpring}, but we also have to take into account of the optical spring for the angular motion. The effect of optomechanical coupling to the angular motion is especially important in pendulum experiments since their natural mechanical frequency of the angular motion is usually equivalent or much smaller than the pendulum mode frequency, which is not the case in cantilever experiments. The angular motion of the mirror creates tilt of the cavity beam axis, and the radiation pressure creates restoring or anti-restoring force depending on the cavity geometry. This effect is called Sidles-Sigg effect~\cite{Sidles-Sigg}.

\begin{figure}
  \begin{center}
    \resizebox{0.7\columnwidth}{!}{\includegraphics{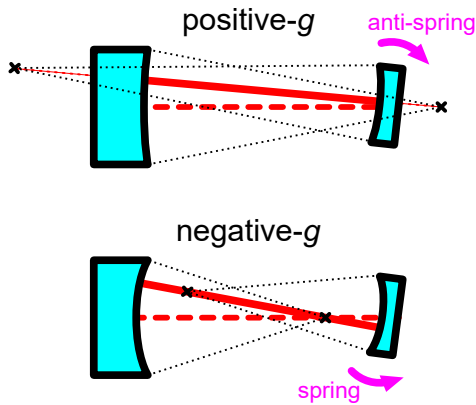}}
  \end{center}
  \caption{The cavity axis change from the tilt of a movable mirror in a Fabry-P\'{e}rot cavity. The radiation pressure torque depicted in pink arrows acts as an anti-spring in a positive-$g$ cavity (top), while it acts as a spring in a negative-$g$ cavity (bottom). Cross mark denote the centers of curvature of the mirrors.}
  \label{SidlesSigg}
\end{figure}

Figure.~\ref{SidlesSigg} depicts the Sidles-Sigg effect for a Fabry-P\'{e}rot cavity consist of a fixed mirror and a movable mirror with different cavity geometries. $g$-factor of a mirror is defined by
\begin{equation}
 g_{i} = 1- \frac{L}{R_i}, \qquad (i = 1,\, 2)
\end{equation}
where $R_i$ is the radius of curvature. Hereafter, we denote fixed mirror by 1 and movable mirror by 2. The cavity stability criterion requires $0 \le g_{\rm 1} g_{\rm 2} \le 1$. The optical torsional stiffness to the movable mirror is given by
\begin{equation}
 k_{\opt}^{\tor} = - \frac{2 \Pcirc L}{c} \frac{g_{\rm 2}}{1-g_{\rm 1} g_{\rm 2}} .
\end{equation}
It is therefore necessary to make a pendulum with high angular mechanical stiffness or to make $g_{\rm 2}$ negative. Torsionally stiff pendulum can be realized by using a thick wire or by suspending the mirror with multiple wires, as was done in Ref.~\cite{Corbitt2007,Neben2012,Kelley2015}, but it requires a careful fabrication to assure symmetric structure, and thicker wire generally result in a larger dissipation, as discussed in the previous section. Negative-$g$ mirror requires a concave mirror, and the fabrication of such a milligram scale mirrors is technically challenging at this point.

The other solution would be to attach actuators, as is done in gravitational wave detectors, but it is also challenging for milligram scale mirrors. Note that in a Fabry-P\'{e}rot cavity consists of two movable mirrors, antisymmetric angular motion, which moves the centers of curvature in the same direction, has negative optical stiffness even with negative-$g$ mirrors. Feedback control of such {\it soft mode} is important for operating the detector with high circulating power~\cite{Lisa2010,AsoKAGRA}.

The pendulum experiment by a group in Tokyo using a 20~mg flat mirror suffered from this torsional anti-spring effect~\cite{Sakata2010}. They have developed a new method to feedback control the angular motion of the other mirror to fix the beam position on the 20~mg mirror, and demonstrated the reduction of the anti-spring effect~\cite{Enomoto2016,Nagano2016}. Matsumoto \etal instead developed a triangular cavity to modify the optical torsional stiffness~\cite{Matsumoto2014,Matsumoto2015}. For an angular motion about the axis perpendicular to the plane of the triangle, radiation pressure torque acts as a spring, even if the mirror is positive-$g$. This is not the case for the angular motion about the axis parallel to the plane of the triangle. The angular motion about the vertical axis (pitch) is often mechanically stiffer than the angular motion about the horizontal axis (yaw), when the mirror is suspended with a single wire attached on the top of the mirror. Therefore, the horizontal triangular cavity is useful for trapping the yaw motion.

\subsection{Torsion pendulum}
Torsion pendulums have been used as highly sensitive force sensors in a variety of applications throughout the history of precision experimental physics~\cite{TorsionReview1,TorsionReview2}. Traditionally, the readout of their angular motion has been done with optical levers, while a very few experiments use Michelson interferometers to read out the differential motion of each edge of the mass~\cite{Ishidoshiro2011}. Although their operating frequencies are usually much higher than suspended torsion pendulums, experiments have been also done at a few hundred milligram scales exploiting microfabricated torsion oscillators. The oscillator made of silicon recorded a $Q$ factor of $\sim 10^5$ and a torque sensitivity of $4 \times 10^{-16}~{\rm Nm /\sqrt{Hz}}$ at the mechanical resonance around 6~kHz~\cite{Haiberger2007}. The oscillator made of tungsten recorded a $Q$ factor of $2.5 \times 10^4$ at the mechanical resonance of 1.2~kHz, and was used for the test of the inverse-square law of gravity at submillimeter range~\cite{Long2003}.

In recent years, several proposals to combine cavity optomechanics with torsion pendulum to enhance the force sensitivity have emerged~\cite{Wang2009,Tan2015}. The experiment by Mueller \etal was one of the first experiments to combine an optical cavity with a torsion pendulum. They reported the observation of optomechanical multistability using a 0.2-g torsion pendulum read out with an optical lever~\cite{MuellerMultistability}. The mass of the torsion pendulum was a gold-coated glass plate, and the torsion pendulum had a thermal noise limited force sensitivity at $3 \times 10^{-12}~{\rm N/\sqrt{Hz}}$ level at 20--100~mHz~\cite{Mueller2008}.

Very recently, Komori \etal reported torque sensing of a 10-mg torsion pendulum with two optical cavities on each edge of a fused silica bar mirror~\cite{Komori2020}. Optical cavities are employed for both optical trapping and readout of the torsional motion. They have successfully increased the torsional resonant frequency from its natural value of 90~mHz to 1~kHz with optical springs, and recorded a torque sensitivity of $2 \times 10^{-17}~{\rm N m/\sqrt{Hz}}$, or, equivalently, a force sensitivity of $3 \times 10^{-15}~{\rm N /\sqrt{Hz}}$ at around 100~Hz.

\begin{figure}
  \begin{center}
    \resizebox{0.9\columnwidth}{!}{\includegraphics{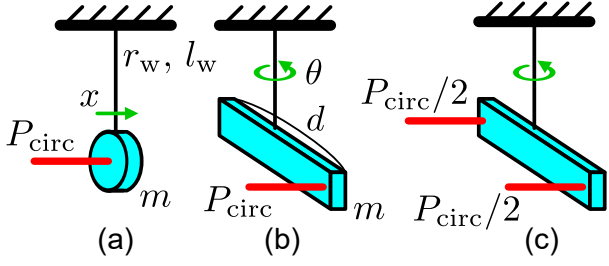}}
  \end{center}
  \caption{Schematic of (a) a longitudinal displacement sensing of a simple pendulum and (b) a torsional motion sensing of a torsion pendulum compared in the main text. The laser power impinging on the mirror $\Pcirc$, the mass of the mirror $m$, the wire radius $r_{\rm w}$ and the length $l_{\rm w}$ are the same for each setup. Torsional motion sensing in (b) and (c) are ideally equivalent in terms of quantum noise and suspension thermal noise.}
  \label{CompareTorsion}
\end{figure}

Let us compare the force sensitivity of a torsion pendulum with a simple pendulum that has the same mass suspended with the same wire, as shown in Fig.~\ref{CompareTorsion}. The displacement of the pendulum is read out by a cavity with circulating power of $\Pcirc$ at the center of the mass, while the torsion of the pendulum is read out by a cavity with the same parameters at the edge of the mass. Note that, in practical experiments, it is better to do differential sensing of the torsional mode using two cavities to avoid mixing of the longitudinal pendulum mode. In terms of quantum noise and suspension thermal noise, the torsion sensing with one cavity and the torsion sensing with two cavities at half the power are ideally equivalent (Fig.~\ref{CompareTorsion}(b) and (c)).

Discussions we have done so far can be easily applied to torsion pendulums by replacing the force $F$ and displacement $x$ with the torque $\tau$ and angle $\theta$, and the mass $m$ with the moment of inertia $I=amd^2$, where $d$ is the length of the torsion bar and a factor $a$ depends on the mass distribution along the bar. When the mass is uniformly distributed along $d$, $a=1/12$, and $a$ takes the largest value of $1/4$ when the mass is concentrated at the edges. The quantum noise in angle also follows the equation similar to \shiki{quantumnoise}, but with $G_{\rm tor}=(d/2)G$ to account for the coupling of angular displacement to the optical frequency shift dependence on $d$. The quantum noise for the angular motion will be
\begin{equation}
 S_{\rm qn}^{\theta}(\omega) = \frac{\theta_{\rm SQL}^2}{2} \left( \frac{1}{\mathcal{K}^{\rm tor}} + \mathcal{K}^{\rm tor} \right),
\end{equation}
with
\begin{equation}
 \theta_{\rm SQL} = \sqrt{2 \hbar | \chi_{\rm I}(\omega) |}
\end{equation}
and
\begin{eqnarray} \label{Eq:Ktor}
 \mathcal{K}^{\rm tor} &=& \frac{4 \hbar G_{\rm tor}^2 \ncirc |\chi_{\rm I}(\omega)|}{\kappa} \frac{1}{1+\left( \omega/\kappa \right)^2} .
\end{eqnarray}
Here, $\chi_{\rm I} (\omega)$ is the mechanical susceptibility for the torsional mode:
\begin{equation}
 \chi_{\rm I} (\omega) = \frac{\theta(\omega)}{\tau(\omega)} = \frac{1}{I [\omegam^2 - \omega^2 + i \gamma_\m \omega]} .
\end{equation}

Quantum cooperativity for this torsion pendulum can be written as the ratio of torque noise from quantum radiation pressure noise and thermal noise,
\begin{equation}
 \Cooperativity_{\rm qu}^{\rm tor} \equiv \frac{S_{\rm rad}^{\tau}}{S_{\rm th}^{\tau}} = \frac{1}{4a} \frac{S_{\rm rad}^{F}}{4 \kB T_{\rm th} m \gamma_{\rm m}^{\rm tor}} .
\end{equation}
Since $a<1/4$, the coefficient $1/(4a)$ tells that the effective mass of a torsion pendulum is smaller, and the quantum cooperativity is larger compared with a simple pendulum that has the same energy damping rate $\gamma_{\rm m}$. The standard quantum limit reaching frequency for the torsion pendulum with $\omegam \ll \omega \ll \kappa$ will be
\begin{equation}
 \omega_{\rm SQL}^{\rm tor} = \sqrt{\frac{S_{\rm rad}^{\tau}}{\hbar I}} = \sqrt{\frac{S_{\rm rad}^{F}}{4a \hbar m}} .
\end{equation}
Compared with $\omega_{\rm SQL}$ of a simple pendulum given in \shiki{Eq:SQLtouch}, it is higher by a factor of $1/\sqrt{4a}$. Higher $\omega_{\rm SQL}$ is advantageous when the thermal noise follows structure damping model, since $\gamma_{\rm m}$ is reduced with frequency.

The drawback of using torsion pendulums is that there are no gravitational dissipation dilution for the torsional mode. The mechanical spring constant of a torsion pendulum suspended with a wire made of an isotropic material is given by
\begin{equation}
 K_{\rm m}^{\rm tor} = \frac{\pi E_{\rm w} r_{\rm w}^4}{4(1+\nu_{\rm w})l_{\rm w}} (1+i \phi_{\rm el}) ,
\end{equation}
where $\nu_{\rm w}$ is the Poisson's ratio of the wire. Assuming that $\phi_{\rm el}$ of the wire is the same for the pendulum mode and the torsional mode, and that dissipation of the system is limited by the structural damping of the wire, the ratio of the damping rate for the pendulum mode and the torsional mode is given by
\begin{equation}
 \frac{\gamma_{\rm m}^{\rm tor}}{\gamma_{\rm m}^{\rm pend}} = \frac{l_{\rm w} r_{\rm w}^2}{a(1+\nu_{\rm w})d^2} \sqrt{\frac{\pi E_{\rm w}}{m \ggrav}} .
\end{equation}
Using the tensile strength given in \shiki{Eq:tensile}, this reduces to
\begin{equation}
 \frac{\gamma_{\rm m}^{\rm tor}}{\gamma_{\rm m}^{\rm pend}} = \frac{l_{\rm w} s_{\rm w}}{a(1+\nu_{\rm w}) H_{\rm w} d^2} \sqrt{\frac{m \ggrav E_{\rm w}}{\pi}} .
\end{equation}
Therefore, when heavy mass is suspended with a long wire, pendulum mode tends to have lower thermal noise because of gravitational dilution. However, if the tensile strength of the wire is high and the wire can be made thin, torsional mode can have a very low mechanical resonant frequency and can realize lower thermal noise. For example, for a $m=10$~mg and $d=10$~mm bar suspended with a silica fiber of $r_{\rm w}=1.5~\mu$m and $l_{\rm w}=10$~mm, $\gamma_{\rm m}^{\rm tor}/\gamma_{\rm m}^{\rm pend}$ is about 0.1, and the thermal noise in amplitude is 3 times smaller for torsional mode. The torsion pendulum experiment by Komori \etal used a single strand of carbon fiber to suspend a bar mirror~\cite{Komori2020}, since carbon fiber has high tensile strength, low shear modulus and low density~\cite{CarbonFiber}. The thinnest carbon fiber available off the shelf is 5--7~$\mu$m in diameter, and technical developments are necessary to fabricate thinner carbon fibers. Larger $d$ is also effective in reducing the thermal noise of torsion pendulum, if $m$ stays constant. In reality, larger $d$ will make the bending mode frequency of the bar lower, which could contaminate the measurement, as bending mode frequency scales with $1/d^2$.

It is worth noting that the surface loss, clamping loss and bonding loss also exist in torsion pendulums, but thermo-elastic damping is very small. This is because the volumetric change of the wire is not involved to the first order approximation in torsional mode. As mentioned earlier, it is also important to note that the decoupling of longitudinal mode should be taken into account to design torsion pendulum experiments. One solution is to use a doubly suspended torsion pendulum to increase the mechanical resonant frequency of the pendulum mode, as was done in Ref.~\cite{Mueller2008}. The other solution is to use two cavities to perform a differential sensing of the torsional mode, as was done in Ref.~\cite{Komori2020} and depicted in Fig.~\ref{CompareTorsion}(c). It is necessary to make two cavities as identical as possible to increase the common mode rejection of the pendulum mode. The common mode rejection must be better than $\sqrt{\gamma_{\rm m}^{\rm tor}/\gamma_{\rm m}^{\rm pend}}$ in amplitude to make the mixing of the thermal noise from the pendulum mode negligible.

Lastly, we comment on the comparison of quantum noise between the angular motion sensing with an optical lever and an optical cavity. The standard quantum limit for optical levers has been studied in Ref.~\cite{EnomotoOplev}. When the Gouy phase shift of the beam from the mirror and the detection point is $\pi/2$, the $\mathcal{K}$ factor for the optical lever is given by
\begin{equation}
  \mathcal{K}_{\rm oplev}^{\rm tor} = \frac{2 \omega_{\rm L} P_{\rm oplev} w^2 |\chi_{\rm I}(\omega)|}{c^2} ,
\end{equation}
where $P_{\rm oplev}$ is the power of the optical lever beam and $w$ is the beam radius at the mirror. If we compare this with \shiki{Eq:Ktor}, it follows that when $P_{\rm oplev} = (2 \mathcal{F}/\pi) \Pcirc$ and $w=d$, the quantum noise for the optical lever and the optical cavity are equal. For a bar shaped mirror, it is difficult to make the size of the beam equivalent to the length of the bar. For a disk mirror with small $m$, the laser power required to reach the standard quantum limit is relaxed and the optical lever could be a good torsion sensor.

\subsection{Optical levitation}
\begin{figure}
  \begin{center}
    \resizebox{0.8\columnwidth}{!}{\includegraphics{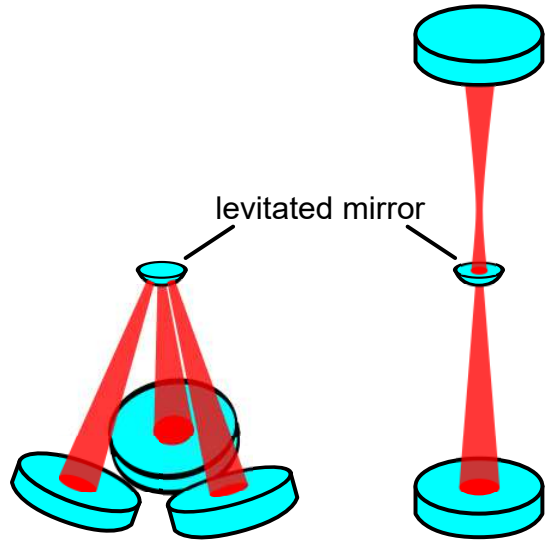}}
  \end{center}
  \caption{Two configurations for optical levitation of a mirror: tripod configuration (left) and sandwich configuration (right).}
  \label{OpticalLevitation}
\end{figure}

Most of mechanical oscillators are supported by some mechanical structures and it is hard to avoid the dissipation from the support. Optical trap of oscillators with an optical spring is a powerful tool to dilute the mechanical loss, and its extreme example is to levitate an object by the optical field alone. Conventional method for the optical levitation is to use optical tweezer originally developed by Ashkin to levitate nanoscale to microscale particles that are transparent to laser beam~\cite{Ashkin1971,Grier2003,Chang2010,Li2011}. There has been a great interest in probing quantum physics with levitated optomechanical systems, and there has been a rapid progress in both experimental and theoretical studies in recent years. Very recently, several groups even reported the ground state cooling of a levitated nanoparticle~\cite{Tebbenjohanns2020,Delic2020}.

Although optical tweezers are commonly used in optical levitation experiments, there is a limitation that the mass of the particle can be only up to nanogram scales. This is because optical tweezers use optical gradient force in a highly focused laser beam, and the size of the particle has to be smaller than the size of the beam~\cite{Grier2003}. Recently, optical levitation of a highly reflective mirror with vertical Fabry-P{\'e}rot cavities have been proposed and studied~\cite{Guccione2013,Michimura2017,MarcoHo2019,Lecamwasam2019}. The levitated mirror act as an upper mirror of the cavity and the mirror is supported by the radiation pressure of the intracavity field. Because of this scattering-free configuration, more massive mirrors can be levitated and optomechanical coupling can be maximized.

To levitate a mirror, the motion of the mirror in all the six degrees of freedom has to be trapped by the optical field and gravity. The vertical motion of the mirror which changes the cavity length can be trapped by an optical spring. The rotational motion around the center of curvature of the levitated mirror do not change the cavity axis, and if the mirror is convex downward, gravity act as a restoring torque for these rotational motions. Trapping of the horizontal motion requires some arrangements of the optical cavities, as shown in Fig.~\ref{OpticalLevitation}. Guccione \etal proposed the use of tripod configuration to couple the optical spring also to the horizontal motion, and proposed to levitate 0.3~mg fused silica mirror~\cite{Guccione2013}. Michimura \etal instead proposed to use two Fabry-P{\'e}rot cavities vertically aligned to sandwich the levitated mirror, and showed that reaching the standard quantum limit of 0.2~mg mirror is feasible with this configuration~\cite{Michimura2017}. Since the levitated mirror is convex downward, the lower cavity has to be positive-$g$ cavity, and the radiation pressure act as an anti-spring as discussed in Sec.~\ref{SEC:Instability}. By forming an upper cavity with negative-$g$, trapping of the horizontal motion can be possible. The trapping of the horizontal motion with the sandwich configuration is recently demonstrated using a torsion pendulum as a force sensor~\cite{Kawasaki2020}.

\begin{table*}
\caption{Suspension parameters of examples of milligram to gram scale experiments. Most of the experiments focuses on measuring the pendulum mode or the longitudinal mode, while the experiments by Mueller \etal~\cite{Mueller2008} and Komori \etal~\cite{Komori2020} focuses on measuring the torsional mode. The dimensions and material of the wire are written together with bonding material or method in parenthesis. $f_\m$ and $Q_\m$ are natural mechanical resonant frequency and natural mechanical $Q$ factor, respectively. Unless otherwise noted, the mass is suspended with a single wire. Optomechanical micromirror experiments at similar scales not discussed in the main text are also listed as comparison.} \label{tab:experiments}
\begin{tabular}{lrllrr}
\hline\noalign{\smallskip}
 & Mass $m$ & Size of mass & Wire (Bonding) & $f_\m$ (Hz) & $Q_\m$ \tabularnewline
\noalign{\smallskip}\hline\noalign{\smallskip}
Arcizet~(2006)~\cite{Arcizet2006} & 0.19~mg & 1~mm$\times$1~mm$\times$60~$\mu$m &  Doubly-clamped silicon beam & 8.14~k & $1 \times 10^4$ \tabularnewline
Michimura~(2017)~\cite{Michimura2017} & 0.2~mg & 0.7~mm dia., t 0.23~mm & \multicolumn{3}{l}{Optical levitation with upper and lower cavities} \tabularnewline

Pontin~(2018)~\cite{Pontin2018} & 0.25~mg & 0.4~mm dia., t $\sim70~\mu$m & Silicon micromirror at 4.9~K & 170~k & $1.1 \times 10^6$ \tabularnewline

Guccione~(2013)~\cite{Guccione2013} & 0.3~mg & 2~mm dia. & \multicolumn{3}{l}{Optical levitation with tripod cavities} \tabularnewline
Matsumoto~(2015)~\cite{Matsumoto2015} & 5~mg & 2~mm dia., t 0.2~mm & $2r_{\rm w}=3~\mu{\rm m}$, $l_{\rm w}=5~{\rm cm}$ & 2.14 & $1.8 \times 10^2$ \tabularnewline
 & & & Tungsten (Epoxy) & & \tabularnewline
Matsumoto~(2019)~\cite{Matsumoto2019} & 7~mg & 3~mm dia., t 0.5~mm & $2r_{\rm w}=1~\mu{\rm m}$, $l_{\rm w}=1~{\rm cm}$ & 4.4 & $1 \times 10^5$ \tabularnewline
 & & & Silica (Epoxy) & & \tabularnewline
Cata{\~n}o-Lopez~(2019)~\cite{Seth2019} & 7~mg & 3~mm dia., t 0.5~mm & $2r_{\rm w}=1~\mu{\rm m}$, $l_{\rm w}=5~{\rm cm}$ & 2.2 & $2 \times 10^6$ \tabularnewline
 & & &  Silica (Monolithic) & & \tabularnewline
Komori~(2020)~\cite{Komori2020} & 10~mg & 15~mm$\times$1.5~mm$\times$0.2~mm & $2r_{\rm w}=6~\mu{\rm m}$, $l_{\rm w}=2.2~{\rm cm}$ & 0.09 & $2.6 \times 10^3$ \tabularnewline
\quad Torsion & & & Carbon (Epoxy) & & \tabularnewline
Sakata~(2010)~\cite{Sakata2010} & 20~mg & 3~mm dia., t 1.5~mm & $2r_{\rm w}=10~\mu{\rm m}$, $l_{\rm w}=1~{\rm cm}$ & 3.7 & N.A. \tabularnewline
 & & &  Silica (Epoxy) & & \tabularnewline
Mueller~(2008)~\cite{Mueller2008} & 0.2~g & 50~mm$\times$10~mm$\times$0.15~mm & $2r_{\rm w}=25~\mu{\rm m}$, $l_{\rm w}=15~{\rm cm}$ & 0.36 & $2.3 \times 10^3$ \tabularnewline
\quad Torsion & & &  Two tungsten, doubly-clamped & & \tabularnewline
Altin~(2017)~\cite{Altin2017} & 0.3~g & 6.35~mm dia. & 100~$\mu$m thick silicon cantilever & 165 & $5.5 \times 10^4$ \tabularnewline
Mow-Lowry~(2008)~\cite{Mow-Lowry2008} & 0.69~g & 7~mm dia., t 1~mm & 300~$\mu$m thick niobium cantilever & 84.8 & $4.5 \times 10^4$ \tabularnewline
Corbitt~(2007)~\cite{Corbitt2007} & 1~g & 12~mm dia., t 3~mm & $2r_{\rm w}=300~\mu{\rm m}$ & 12.7 & $2.0 \times 10^4$ \tabularnewline
 & & & Two optical fibers (Epoxy)& & \tabularnewline
Neben~(2012)~\cite{Neben2012} & 1~g & 12~mm dia., t 3~mm & $2r_{\rm w}=150$--$3000~\mu{\rm m}$, $l_{\rm w}=4~{\rm cm}$ & 10 & $1 \times 10^6$ \tabularnewline
 & & & Two silica fibers~(Epoxy) & & \tabularnewline
\noalign{\smallskip}\hline
\end{tabular}
\end{table*}

The laser power required to levitate a mirror can be easiliy calculated with
\begin{equation}
 \Pcirc^{\rm lev} = \frac{m \ggrav c}{2},
\end{equation}
where $\Pcirc^{\rm lev}$ is the sum of all the power impinging on the levitated mirror projected to the vertical axis. For a 1-mg mirror, $\Pcirc^{\rm lev} \simeq 1.5$~kW. For simplicity, if we assume $\Pcirc^{\rm lev}$ comes from the lower cavity which also reads out the vertical displacement of the mirror, the standard quantum limit reaching frequency given in \shiki{Eq:SQLtouch} reduces to
\begin{equation}
 \omega_{\rm SQL} = \sqrt{\frac{16 \Finesse \ggrav}{\lambda}} ,
\end{equation}
which is independent of $m$. Since the quantum radiation pressure effects reduces above the cavity pole frequency $\kappa$, it is better to have $\omega_{\rm SQL} \ll \kappa$, and this condition reduces to
\begin{equation} \label{Eq:FinesseLevitation}
  \Finesse \ll \left( \frac{\pi^2 c^2 \lambda}{64 L^2 \ggrav} \right)^{\frac{1}{3}} .
\end{equation}
Therefore, although high circulating power is necessary to levitate a massive mirror, finesse of the cavity cannot be very high to reach the standard quantum limit. For a cavity with length $L=10$~cm and laser wavelength $\lambda=1064$~nm, the finesse requirement is $ \Finesse \ll 2.5 \times 10^3$, and for $\Finesse = 100$, $\omega_{\rm SQL} \simeq 2 \pi \times 19~$kHz, which is relatively high compared with other milligram scale pendulum experiments, and this discussion is independent of $m$. One way to circumvent this power constraint in choosing $\omega_{\rm SQL}$ is to read out the other degrees of freedom rather than reading out the vertical displacement of the levitated mirror.

High circulating power to levitate the mirror also leads to the heating of the mirror, which deforms and changes the refractive index of the mirror substrate and its high reflective coating. Such photothermal effects have to be correctly taken into account to design a stable optical levitation, and recent studies show that careful engineering of photothermal effects can stabilize a system~\cite{Kelley2015,Altin2017,Lecamwasam2019}. Fabrication of a low absorption milligram scale convex mirror is still technically challenging, and this is one of the main challenges towards the realization of the optical levitation of a mirror. Recently, Ilic \etal proposed adding nanophotonic structures on the substrate surface to enhance the optical levitation~\cite{PClevitation}.

Despite the technical challenges, optical levitation is attractive in that all the dissipation mechanisms associated with wires, including the coupling of the wire violin modes and other unwanted eigenmodes, are absent. Thermal noise from the residual gas damping and thermal noise from the mechanical loss of the mirror itself are still present in the optical levitation. If dissipation mechanisms associated with wires can be made small enough compared with residual gas damping, the $Q$ factor of pendulums can be as large as that of optically levitated mirrors.

Using the parameters listed in Ref.~\cite{Michimura2017}, quantum cooperativity for the 0.2~mg optically levitated mirror at $10^{-5}$~Pa will be $2 \times 10^3$. With $f_{\rm eff}=340$~Hz and $\gamma_{\rm m}^{\rm gas}=7 \times 10^{-8}$~Hz, $f_{\rm eff} Q_{\rm eff}=1 \times 10^{13}$ can also be achieved. Once the optical levitation of a mirror is realized, meeting the criteria discussed in Sec.~\ref{Sec:Criteria} is feasible due to high circulating power and low dissipation.

\section{Summary}
Optomechanical experiments at milligram scales are receiving increased attention in probing the boundary between classical and quantum mechanics. In this paper, we reviewed the key concepts of cavity optomechanics and summarized the criterions for reaching the quantum regime, especially for the case when the mechanical resonant frequency is much below the linewidth of the cavity. We showed that making quantum cooperativity larger than unity is equivalent to making quantum radiation pressure noise larger than the thermal noise. We also showed that optical dilution of mechanical dissipation is especially important for pendulum experiments with suspension thermal noise which follow the structure damping model.

Realistic comparison between pendulum, torsion pendulum and optical levitation experiments at milligram scales are also presented. We derived achievable $Q$ factor for pendulums, and showed that a wire with high tensile strength and low density is necessary to achieve a high $Q$ pendulum. We then derived the criteria for preparing more sensitive force sensor with torsion pendulums compared with longitudinal displacement sensing. Advantages and constraints of the optical levitation of a mirror are discussed, and future prospects in circumventing these constraints are presented. Milligram scale optomechanical systems can be used for sensing force from subtle effects, such as microgravity, decoherence mechanisms and dark matter. Bringing milligram scale oscillators into quantum regime is within our experimental reach, and quantum tests of gravity theories would be possible in the next few decades.

We thank Nobuyuki Matsumoto, Yutaro Enomoto, Koji Nagano, Ching Pin Ooi and Takuya Kawasaki for insightful discussions. This work was supported by JSPS KAKENHI Grant Numbers 18H01224, 18K18763, and 201960096, JST CREST Grant Number JPMJCR1873, and MEXT Q-LEAP Grant Number JPMXS0118070351.


\end{document}